\documentclass[showpacs,amsmath,amssymb,aps,preprint,groupedaddress]{revtex4-1}
\usepackage{amsmath}    
\usepackage{amsfonts}  
\usepackage[latin1]{inputenc}

\begin{document}
\newtheorem{theorem}{Theorem}[section]
\newtheorem{definition}{Definition}[section]
\newcommand{\connection}[3]{\stackrel{#1}{\Gamma}^{\mathrm{\raisebox{-0.1cm}{$_{#2}$}}}_{\ #3}}
\newcommand{\covariant}[2]{\!\stackrel{#1}{\nabla}_{#2}\!}
\newcommand{\covariantdown}[2]{\!\mathop{\stackrel{#1}{\nabla}}_{#2}\!}
\newcommand{\riemann}[3]{\stackrel{#1}{R}^{\mathrm{\raisebox{-0.1cm}{$_{#2}$}}}_{\  #3}}

\title{Dirac equation in non-Riemannian geometries}
\author{J. B. Formiga}
 \affiliation{Centro de Ciências da Natureza, Universidade Estadual do Piauí, C. Postal 381, 64002-150 Teresina, Piauí, Brazil}
 \email{jansen.formiga@uespi.br}   
\author{C. Romero}
\affiliation{Universidade Federal da Paraíba, Departamento de Física, C. Postal 5008, 58051-970 João Pessoa, Pb, Brazil}
\date{\today}

\begin{abstract}
We present the Dirac equation in a geometry with torsion and non-metricity balancing generality and simplicity as much as possible. In doing so, we use the vielbein formalism and the Clifford algebra. We also use an index-free formalism which allows us to construct objects that are totally invariant. It turns out that the previous apparatuses not only make possible a simple deduction of the Dirac equation but also allow us to exhibit some details that is generally obscure in the literature.
\end{abstract}

\maketitle

\section{Introduction}
The Dirac equation is one of the most important achievement of modern physics. It is in good agreement with experiments and is the core of many different theories that try to explain the laws of physics. However, this equation was first obtained in the Minkowski spacetime with Cartesian coordinates and in an inertial frame. Its version in a non-inertial frame, or equivalently,  in a curved space-time, has been widely used in the literature \cite{Kiefer,Gasperinibook,Nakahara,Brill:1066,Hammond:2002rm,Poplawski}, and seems to agree with experiments\cite{Gronwald9602013}. Despite the arbitrariness in the definition of the spin connection \cite{Crawford:2003hy}, one usually presents it in a simple form \cite{Kiefer,Gasperinibook,Nakahara,Brill:1066,Blagojevic:2002du,Hammond:2002rm,Poplawski,Fariborz}. For beginners the Dirac equation in a curved spacetime may look like very intriguing. This is even worse when dealing with non-Riemannian theories such as the Eintein-Cartan (see e.g., Ref.~\cite{Hammond:2002rm}). Our goal is to present some results that we hope will help any reader to have a better understanding of the Dirac equation in curved spacetimes. In addition, we also hope to give a useful guide to the literature on the subject. The reader may also check Ref. \cite{Adak:2002pq} for a completely different approach of the Dirac equation in non-riemannian geometries. There, the authors use the minimal coupling procedure applied to the field equation (MCPE), which yields a different Dirac equation.  

In our notation Greek letters represent coordinate indices, capital Latin letters indicate the vielbein indices, which appears between parentheses when using numbers; for example, $e_A=(e_{(0)},e_{(1)},...)$. We leave the small Latin letters to the ``spin indices''. In this article we avoid the standard spinor formalism\cite{Penrose1,Naberspin}, since it uses much more indices than we will need here. We will even make use of an index-free notation whenever it is convenient. 

This paper is organized as follows. In Sec.~\ref{02032011a} we present the vielbein formalism\footnote{The vielbein formalism generalizes the vierbein formalism (tetrad formalism) to a $n$-dimensional spacetime.}. In Sec.~\ref{02032011b} we give an index-free definition of the Dirac spinor, while in Sec~\ref{02032011c} we introduce the Dirac matrices. Sec.~\ref{gdac12092010} is devoted to the Clifford algebra. From Sec.~\ref{05032011a} to \ref{05032011b} we focus on obtaining the Dirac equation, then, in Sec. \ref{02032011d}, we conclude with some final remarks.

\section{\label{02032011a} Vielbein}
We call ``vielbein''  a set of one-forms $\{\theta^A \}$ such that the metric written in terms of it coincides with the Minkowski metric, i.e., $g=\eta_{AB}\theta^A\otimes\theta^B=(\theta^{(0)})^2-(\theta^{(1)})^2 -\ldots -(\theta^{(n-1)})^2$, where $n$ is the spacetime dimension and we are using Einstein summation convention. Its dual basis\footnote{Saying that a basis of one-forms $\omega^A$ is dual to a basis of vectors $v_A$ means that $\omega^B(v_A)=\delta^B_A$.} is a set of vector fields, which we denote by $\{e_A\}$. Indeed, the term ``vielbein'' will be used for both sets. In the coordinates bases $\{dx^{\mu}\}$ and $\{\partial_{\mu} \}$, we have  $\theta^A=e^A_{\ \ \mu}dx^{\mu}$ and $e_A=e_A^{\ \ \mu}\partial_{\mu}$, where $e^A_{\ \ \mu}$ and $e_A^{\ \ \mu}$ are the components of $\theta^A$ and $e_A$ in these bases, respectively. From the properties of the vielbein, it follows $e_B^{\ \ \mu}e^A_{\ \ \mu}=\delta^A_B$ and $e_A^{\ \ \mu}e^A_{\ \ \nu}=\delta^{\mu}_{\nu}$.

The above definition does not fix the vielbein, since there is still an infinite number of bases we can take. Nonetheless, all of them are related to each other by a local Lorentz transformation $\Lambda$. For example, if you choose a vielbein $\theta^A$ and someone else take other, say $\overline{\theta}^B$, then $\overline{\theta}^B=\Lambda^B_{\ \ A}\theta^A$. When we pass from one basis to another some quantities change, one of them is the Dirac spinor. How this spinor changes is one of the subjects discussed here.

\section{\label{02032011b} Dirac spinor}
In the standard approach, spinors are invariant under coordinate transformation but changes in a well-defined way when one passes from one vielbein to another. So, we can construct a basis that changes exactly in the opposite way, and then compensate the change of the spinor. By writing the spinor in this basis, we manage to create a mapping that takes values in the spacetime manifold $M$ and leads to a complex number. Let us start with the definition of the spin basis.
\begin{definition} \label{basisspinordefinition}
Let $p$ be a point in a n-dimensional manifold $M$, $\xi_a:M \to \mathbb{C}^n$ a set of mappings defined at $p$, and $S$ is a representation of the proper Lorentz group $L^{\uparrow}_+$ in the vectorial space $\mathbb{S}$ which is built up from the set of functions $\{\xi_a\}$. We call $\mathbb{S}$ the spin space, where   $\xi_a$ represents its basis. Furthermore, given two bases of $\mathbb{S}$, say $\{\xi'_a \}$ and $\{\xi_a \}$, we demand  
\begin{equation}
\xi'_a=(S^{-1})^c_{\ a}\xi_c. \nonumber
\end{equation} 
\end{definition}
Now we define the spinor.
\begin{definition}\label{spinordefinition}
Let $\xi_a$ be a basis of $\mathbb{S}$ and  $\psi^a:M \to \mathbb{C}^n$ a set of mappings, both defined at $p \in M$. We call spinor the mapping $\tilde{\psi}:M \to \mathbb{C}$ such that $\tilde{\psi}\equiv \psi^a\xi_a$, where, under the transformation $\xi'_a=(S^{-1})^c_{\ a}\xi_c$, we have
\begin{equation}
\psi'^a=S^a_{\ b}\psi^b. \nonumber
\end{equation}
\end{definition}
The definition of the duals of both $\xi_a$ and $\tilde{\psi}$ are straightforward.
\begin{definition}\label{basiscospinordefinition}
Consider $p \in M$ and a linear mapping $\xi^a:M \to \mathbb{C}^n$ defined at $p$. We demand that this mapping be such that
\begin{equation}
\xi^b(\xi_a)\equiv \delta^b_a. \nonumber
\end{equation}
We call the space generated by $\xi^a$ co-spin space and denote it by $\mathbb{S}^{*}$. 
\end{definition}
Since $\xi^a$ is a linear function of $\xi_b$, we have $\xi^a(f\xi_b)=f\xi^a(\xi_b)=f\delta^a_b$, where $f$ is a function of a variable independent of $\xi_b$. It is also clear that $\xi^a$ transforms like
\begin{equation}
\xi'^d=S^d_{\ b}\xi^b \label{13082010b}.
\end{equation}
The name ``co-spin space'' sounds rather tiresome, hence we will eventually stop using it.
\begin{definition} \label{cospinordefinition}
Let $p\in M$, $\xi^a \in \mathbb{S}^*$ and $\psi_a:M \to \mathbb{C}^n$. The mapping $\hat{\psi}:M \to \mathbb{C}$ such that $\hat{\psi}=\psi_a \xi^a$ is called co-spinor.
\end{definition}
From the definitions \ref{basiscospinordefinition} and \ref{cospinordefinition}, it follows $\hat{\psi}(\tilde{\psi})=\psi_a \psi^a$. We can clearly generalize these definitions for more general spinors by using the tensor product. For instance, we can define objects like $\tilde{\Upsilon}=\Upsilon^{\mu a}_{\ \ b}\partial_{\mu}\otimes\xi_a\otimes\xi^b$.

\section{\label{02032011c} Dirac Matrices}
The Dirac matrices in a vielbein basis are constant and obey the ordinary condition
\begin{equation}
\gamma^A \gamma^B+\gamma^B \gamma^A=2n^{AB}\mathbb{I}. \label{seven}
\end{equation}
So we can choose any representation we are familiar with in the Minkowski spacetime to represent $\gamma^A$. On the other hand, the coordinate version of these matrices are not constant and are defined by
\begin{equation}
\gamma^{\mu} \equiv e_A^{\ \ \mu} \gamma^A. \label{eight}
\end{equation}
Note that this definition ensures $\gamma^{\mu} \partial_{\mu}=\gamma^A e_A$. This is a natural definition if we see the previous index ``$\mu$'' as a tensorial one. An example of the contrary is the affine connection $\Gamma^{\lambda}_{\ \mu \nu}$, although not wrong, it would not be natural to do that with these indices; the definition $\Gamma^A_{\ \ B C}\equiv \Gamma^{\lambda}_{\ \mu \nu}e^A_{\ \ \lambda}e_B^{\ \ \mu}e_C^{\ \ \nu}$ would probably lead one to mistake it for the connection in the vielbein basis, which we denote by $\omega^A_{\ \ BC}$.

The definition (\ref{eight}) leads (\ref{seven}) to
\begin{equation}
\gamma^{\mu}(x) \gamma^{\nu}(x)+\gamma^{\nu}(x) \gamma^{\mu}(x)=2g^{\mu \nu} \mathbb{I}, \label{six}
\end{equation}
\subsection{Transformation properties}
\subsubsection{Coordinate transformation}
As was stated before, the matrices $\gamma^A$s are constant. They do not change under any tranformation, however, the $\gamma^{\mu}$s do. From the definition (\ref{eight}) we see that under coordinate transformation the $\gamma^{\mu}$ transforms like $e_A^{\ \ \mu}$ does, that is, like a vector
\begin{equation}
\gamma'^{\mu}=\frac{\partial x'^{\mu}}{\partial x^{\nu}} \gamma^{\nu}. \label{nine}
\end{equation}
\subsubsection{Rotation of the vielbein}
Since the indices $a$ and $b$ in $\gamma^{\mu a}_{\quad b}$ are supposed to be spin indices, under the vielbein rotation $\overline{\theta}^A=\Lambda^A_{\ \ B}\theta^B$ we must have
\begin{equation}
\overline{\gamma}^{ \mu}=S\gamma^{\mu} S^{-1}. \label{24082010c}
\end{equation}
By using the fact that the matrices $\gamma^A$s do not change, one easily verifies that $S^{-1}\gamma^DS=\Lambda^D_{\ \ B}\gamma^B,$ and $S^{-1}\gamma^{\mu}S=\Lambda^{\mu}_{ \ \nu}\gamma^{\nu}$. The transformations (\ref{nine}) and (\ref{24082010c}) make possible the following definition.
\begin{definition} \label{gammadefinition}
Let $\{\partial_{\mu}\} \in T_pM$, $\{\xi_a \} \in \mathbb{S}_p$, $\{\xi^a \} \in \mathbb{S}^*_p$, and  $\gamma^{\mu a}_{\quad b}:M \to \mathbb{C}^{n^3}$. The mapping $\tilde{\gamma}:T_pM\times \mathbb{S}_p\times \mathbb{S}^*_p \to \mathbb{C}$ is defined by 
\begin{equation}
\tilde{\gamma}\equiv \gamma^{\nu a}_{\quad b} \partial_{\nu} \otimes \xi_a \otimes \xi^b. \nonumber
\end{equation}
\end{definition}
Keep in mind that when we write $\gamma^{\mu a}_{\quad b}$ we are making all the indices of $\gamma^{\mu}$ explicit. In addition, notice that $\tilde{\gamma}$ is totally invariant. It is worth mentioning that \ref{gammadefinition} is not a standard definition; perhaps, one will not find it anywhere else. What the reader will probably find frequently in the literature is the one-form $\gamma \equiv \gamma_{\mu} dx^{\mu}$, which is not invariant under rotation of the vielbein.

\section{\label{gdac12092010} Clifford Algebra} 
A set of matrices $\gamma^A$s is  said to obey the Clifford algebra if  $\{\gamma^A,\gamma^B\}=\gamma^A\gamma^B+\gamma^B\gamma^A=2\eta^{AB}$. In fact, the anticommutator need not be equal to $2\eta^{AB}$, it could be also equal to the Kronecker delta; but since we are thinking of the $\gamma^A$s as the Dirac matrices, we will equal it to $2\eta^{AB}$.  In a n-dimensional spacetime the generators of this algebra are \cite{Nakahara}
\begin{eqnarray}
{\bf \Gamma^d_{m}}\equiv \{\mathbb{I},\ \gamma^A,\ \gamma^{A_1} \gamma^{A_2}\ (A_1<A_2),\  \ldots,
\nonumber \\
\gamma^{A_1}\ldots \gamma^{A_k}\ (A_1< \ldots <A_k),\ldots,
 \nonumber \\
  \gamma^{(0)} \ldots \gamma^{(n-1)}\}, \label{23082010a}
\end{eqnarray}
where the indices $A_i$ run from $(0)$ to $(n-1)$, the index $m$ indicates the elements separated by the commas, and $d$ represents a particular element that is between two commas. As an example, we have ${\bf \Gamma^1_{3}}=\gamma^{(0)}\gamma^{(1)}$. It is important to say that we have omitted the coefficients of the generators in the list (\ref{23082010a}), since they will not be important for us. Besides, one can remove the restrictions on the indices indicated in (\ref{23082010a}) by antisymmetrizing them. For instance, instead of using $\gamma^{A_1}\gamma^{A_2}$ with $A_1<A_2$, we can use just $\gamma^{[A}\gamma^{B]}$.

Except for ${\bf \Gamma^d_1}$, the generators (\ref{23082010a}) are traceless. Besides, they all satisfy ${\bf \Gamma^d_{m}\Gamma^d_{m} }=\pm \mathbb{I}$. As a result of these two properties, they are linearly independent. The number of generators in (\ref{23082010a}) is $2^n$, which allows us to expand any matrix $A \in \mathbb{C}^{2^{n/2}\times 2^{n/2}}$ in terms of them for $n$ even. It is important to emphasize here that $\gamma^A \in \mathbb{C}^{2^{n/2}\times 2^{n/2}}$ for $n$ even, but for $n$ odd, one usually takes $\gamma^A \in \mathbb{C}^{2^{(n-1)/2}\times 2^{(n-1)/2}}$. In the latter case the last Dirac matrix is chosen to be proportional to $\gamma^{(0)}\gamma^{(1)}...\gamma^{(n-1)}$. 

\section{\label{05032011a}Spin connection}
\subsection{Definition}
\begin{definition} \label{spinorconnectiondefinition}
Consider two spinors $\chi$, $\psi$, two function $f,g \in C^{\infty}$, and two vectors $V,U\in T_pM$. We call spinor connection the mapping $\nabla:T_pM\times\mathbb{S}_p \to \mathbb{S}_p$ such that:
\begin{eqnarray}
\nabla_{fV+gU} \psi=f\nabla_{V} \psi+g\nabla_{U} \psi, \nonumber \\
\nabla_V (\psi+\chi)=\nabla_{V}\psi+\nabla_{V} \chi, \nonumber\\
\nabla_V(f\psi)=V[f]\psi+f\nabla_V\psi. \nonumber
\end{eqnarray}
We denote the components of the spinor connection by $\Gamma^b_{\ \mu a}\equiv \xi^b(\nabla_{\mu}\xi_a)$, which implies $\Gamma^a_{\ \mu b}=-\xi_ b(\nabla_{\mu}\xi^a)  $.
\end{definition}
 The action of this connection on a more general object such as $\tilde{\gamma} \in T_pM\otimes \mathbb{S}_p\otimes \mathbb{S}^*_p$ can be realized if we demand that $\nabla$ obey the Leibniz rule when acting on the basis, i.e., if $\nabla_V( \partial_{\alpha}\otimes\xi_a\otimes\xi^b)=( \nabla_V\partial_{\alpha})\otimes\xi_a\otimes\xi^b+\partial_{\alpha}\otimes (\nabla_V \xi_a )\otimes\xi^b+\partial_{\alpha}\otimes  \xi_a \otimes (\nabla_V\xi^b )$. Thus, one can easily verify that the components of the covariant derivative of $\tilde{\gamma}$ read
 $\gamma^{\alpha}_{\ \,|\, \mu}=\gamma^{\alpha}_{\ \, ,\, \mu}+ \connection{}{\alpha}{\mu \nu} \gamma^{\nu} + [\Gamma_{\mu}, \gamma^{\alpha}]$.

\subsection{Transformation properties}
Now we show how the spinor connection transforms under local Lorentz transformation, which is defined by $\theta^{\prime A}=\Lambda^A_{\ \ B}\theta^B$, and a coordinate transformation. It is important to emphasize that these transformations are independent of each other, i.e., a coordinate transformation does not imply a local Lorentz one, and  vice versa.

\subsubsection{Local Lorentz transformation}
From the definitions \ref{basisspinordefinition}, \ref{basiscospinordefinition} and \ref{spinorconnectiondefinition}, it follows
\begin{equation}
  \Gamma'_{ \mu}= S\Gamma_{ \mu }S^{-1}-S_{\ , \mu}S^{-1}.
\end{equation}

\subsubsection{Coordinate transformation}
If we perform a coordinate transformation, we will have $\partial'_{\mu}=(\partial x^{\nu}/ \partial x'^{\mu}) \partial_{\nu}$. Therefore,
\begin{equation}
\Gamma'^{a}_{\ \mu b}=\xi^a\left(\covariantdown{}{\mu'}\xi_b \right)=\frac{\partial x^{\nu}}{\partial x'^{\mu}} \xi^a\left(\covariantdown{}{\nu}\xi_b \right)=\frac{\partial x^{\nu}}{\partial x'^{\mu}} \Gamma^a_{\ \nu b}.
\end{equation}
It is clear that this connection transforms like the components of a tensor under coordinate transformations.

\subsection{Affine connection in a vielbein basis}\label{cnbdt}
Since the affine connection is not a tensor, its components in a vielbein basis are not $e^A_{\ \ \lambda}e_B^{\ \ \mu}e_C^{\ \ \nu}\Gamma^{\lambda}_{\ \mu \nu}$. Nevertheless, it is not hard to find them if we evaluate the components directly, that is, if we use $\theta^C \left(\nabla_{e_A}e_B \right)$. Denoting these components by $\omega^C_{\ \ AB}$, we find
\begin{equation}
\omega^C_{\ \ AB}=e^C_{\ \ \lambda}e_{B\ \ ;A}^{\ \ \lambda}, \label{21082010b}
\end{equation}
where $e_{B\ \ ;A}^{\ \ \lambda}\equiv e_A^{\ \ \alpha}\left(e_{B\ \ ,\alpha}^{\ \ \lambda}+e_B^{\ \ \mu}  \Gamma^{\lambda}_{\ \alpha \mu} \right)$. Although $\omega^C_{\ \ AB}$ is just the components of the affine connection, some authors refer to it as {\it spin connection}. As pointed out by Hehl in \cite{Hehl:1994ue}, some authors also impose the additional constraint  $e_{C\ \ ;B}^{\ \ \lambda}-\omega^A_{\ \ BC}e_A^{\ \ \lambda}=0$. However, from (\ref{21082010b}) one easily sees that this is not a constraint,  it is just an identity. Defining $e\equiv e_A^{\ \ \mu} \theta^A\otimes \partial_{\mu}$, we can express this identity as $\covariant{}{e_B}e=0$.

One can easily check that $V^C_{\ \ |A}=V^C_{\ \ ,A}+\omega^C_{\ \ AB}V^B$ represents the components of the covariant derivative of a vector $V$ in the vielbein basis $\{e_A\}$.

\subsection{Explicit form of the Spin connection}
The definition \ref{spinorconnectiondefinition} alone is not sufficient to determine an explicit form for the spin connection. To get this explicit form, we need an additional condition. In fact, we need to add more assumptions than we would like to, as will become clear later on.

The covariant derivative of the metric is generally regarded as zero (Riemannian geometry), but here, we consider a more general geometry and so we define
\begin{equation}
N(W,U,V)\equiv \covariant{}{V}g(W,U), \label{26082010e}
\end{equation}
where $N(W,U,V)$ is known as the non-metricity tensor. The condition (\ref{26082010e}) is sufficient to determine the affine connection. Unfortunately, it does not fix the spin connection. 

If we define the invariant object $\tilde{\chi}\equiv \left[(\gamma_{\mu})^a_{\ c} (\gamma_{\nu})^c_{\ b}+(\gamma_{\nu})^a_{\ c} (\gamma_{\mu})^c_{\ b} \right]dx^{\mu}\otimes dx^{\nu}\otimes \xi_a \otimes \xi^b$, we can write (\ref{six}) in the following invariant form
 \begin{equation}
\tilde{\chi}(V,U,\hat{\omega},\tilde{\alpha})=2\tilde{g}(V,U,\hat{\omega},\tilde{\alpha}), \label{30082010a}
\end{equation}
where $\tilde{g}\equiv g_{\mu \nu} dx^{\mu}\otimes dx^{\nu} \otimes \xi_a\otimes\xi^a$, and $\hat{\omega}=\omega_a\xi^a$, $\tilde{\alpha}=\alpha^a\xi_a$. By deriving (\ref{30082010a}) and performing some algebra, one finds 
\begin{equation}
\{\gamma^{\mu}_{\ |\lambda},\ \gamma^{\nu}\}+\{\gamma^{\mu},\ \gamma^{\nu}_{\ |\lambda}\}=-2N^{\mu \nu}_{\quad \lambda}, \label{30082010d}
\end{equation}
where the $\{, \}$ represents the anticommutator, and $\gamma^{\mu}_{\ |\lambda}$ is the component of the covariant derivative of $\tilde{\gamma}$ (see definition \ref{gammadefinition}). Although (\ref{30082010d}) looks like a strong constrain on $\gamma^{\mu}_{\ |\lambda}$, in fact, it is not. A very general solution of Eq.~(\ref{30082010d}), if not the most, is\cite{Treat:1970ih} $\gamma^{\mu}_{\ |\lambda}=[\beta_{\lambda},\gamma^{\mu}]-\frac{1}{2}N^{\mu}_{\ \nu \lambda}\gamma^{\nu}$, where $\beta_{\lambda} \in \mathbb{C}^{2^\frac{n}{2}\times 2^\frac{n}{2}}$ ($n$ even). The term with the commutator in the previous solution is not needed for consistency because it vanishes in Eq.~(\ref{30082010d}). This leaves us with too many degrees of freedom, much more than we probably need to describe particles. In fact, Eq.~(\ref{30082010d}) does not fix the connection $\Gamma_{\lambda}$. One can check this by verifying that the spin connection  disappears in the left-hand side of Eq.~(\ref{30082010d}). To narrow these degrees of freedom down, further assumptions must be made. 

In order to fix the spin connection we can define an inner product in the spin space, which leads to a kind of metric, and then demand that the spin connection be compatible with this metric. However, this still leaves us with many degrees of freedom \cite{Crawford:2003hy}. One may also assume that the covariant derivative of the matrix that represents the operation of charge conjugation $C_{ab}$ vanishes, which, by itself, does not reduce this arbitrariness satisfactorily either (see e.g., Ref.\cite{Crawford:2003hy}, p. 2955). Another assumption, and perhaps the most common in the literature, is the covariant constancy of the Dirac matrices ($\gamma^{\mu}_{\ |\lambda}=0$), which is compatible with (\ref{30082010d}) only if $N=0$. To take the non-metricity tensor into account, one may choose $\nabla_{\lambda}\gamma^{\mu}=-(1/2)N^{\mu}_{\ \nu \lambda}\gamma^{\nu}$ instead. This condition reduces the connection very much. If we consider it, we will arrive at
\begin{equation}[\Gamma_{\alpha},\gamma^{\lambda}]=-\frac{1}{2}\left(N^{\lambda}_{\ \ A \alpha} +2e_{A \ \ ;\alpha}^{\ \ \lambda}  \right)\gamma^A,   \label{dgammathree}
\end{equation}
which can be reduced to $[\mathbb{B}_{\alpha},\gamma^{\lambda}]=0$ if we take $\Gamma_{\alpha}\equiv \mathbb{B}_{\alpha}-\frac{i}{4}\omega_{A\alpha B}  \sigma^{AB}$, where $\omega_{A\alpha B}\equiv \eta_{AC}e^D_{\ \ \alpha}\omega^C_{\ \ D B}$, and $\sigma^{AB}\equiv (i/2)[\gamma^A,\gamma^B]$. By using Schur's lemma (see, e.g., Ref. \cite{Naberspin}, p. 151), one finds that $\mathbb{B}_{\alpha}=A_{\alpha}\mathbb{I}$. Therefore, we finally have
\begin{eqnarray}
\Gamma_{\alpha}=A_{\alpha}\mathbb{I}-\frac{i}{4}\omega_{A\alpha B}  \sigma^{AB}.
 \label{02082010a}
\end{eqnarray}
The vector $A_{\alpha}$ is sometimes associated with the electromagnetic potential vector, however, this vector couples to all spinors without making any distinction among the different charges; it would be like all the spinors had the same charge. At least in four dimension and with $N=0$, one can get rid of this problem by using $C_{ab|\alpha}=0$  together with the covariant constancy of the Dirac matrices \cite{Crawford:2003hy}. For reasons that will be clarified later, one may also demand that $\gamma^{(0)}\Gamma_{\alpha}^{\dagger} \gamma^{(0)}=-\Gamma_{\alpha}$.

\section{The matrix $S$}
When we change the vielbein we do it by means of a local Lorentz transformation $\Lambda \in L^{\uparrow}_+$. Mathematically speaking, what we do is $\overline{\theta}^A=\Lambda^A_{\ \ B}\theta^A$. In turn, this induces a change in the components of the Dirac spinor $\tilde{\psi}$. This change is realized by a matrix $S$ which is the representation of the Lorentz group $L^{\uparrow}_+$ in the spin space. Although we are considering a non-Riemannian geometry, the procedure to get the explicit form of $S$ is the same as that of Minkowski spacetime and leads to
\begin{eqnarray}
S=e^{-i\lambda_{AB}\bar{\epsilon}^{AB}\mathbb{I}}\ e^{-\frac{i}{4}\sigma_{AB}\bar{\epsilon}^{AB}}, \label{25082010a}
\\
S^{-1}=e^{i\lambda_{AB}\bar{\epsilon}^{AB}\mathbb{I}}\ e^{\frac{i}{4}\sigma_{AB}\bar{\epsilon}^{AB}}, \label{25082010b}
\end{eqnarray}
where we may assume $\lambda_{AB}$ to be a real function, although one generally sets $\lambda_{AB}=0$. Any of these assumptions lead to the property $\gamma^{(0)}S^{\dagger} \gamma^{(0)}=S^{-1}$.
 
\section{Matrix representation}
So far, we have been mixing index notation with matrix one, however, we have not used the matrix notation for either $\tilde{\psi}$ or $\hat{\psi}$ yet. Now it is at least convenient to extend this notation to these spinors. To do so, we consider the transformation properties of their components and their respective basis.

In terms of matrices, the components $\psi^a$ are represented by $\psi=\left( \begin{array}{c} \psi^0 \\ \vdots \\ \psi^n  \end{array} \right)$. As $\psi_a$ transforms like $\psi_a'=(S^a_{\ b})^{-1}\psi_a$, we relate it to $\overline{\psi}=\psi^{\dagger}\gamma^{(0)}$. The representation of $\xi^a$ and $\xi_a$ is analogous, that is, $\xi=\left( \begin{array}{c} \xi^0 \\ \vdots \\ \xi^n  \end{array} \right) $ for $\xi^a$, and $\overline{\xi}=\xi^{\dagger}\gamma^{(0)}$ for $\xi_a$. Thus, it follows $\tilde{\psi}=\overline{\xi} \psi$ and $\hat{\psi}=\overline{\psi} \xi$. 

In the matrix notation the components of the covariant derivative of $\psi$ read
\begin{equation}
\psi_{|\mu}=\psi_{ ,\mu}+\Gamma_{\mu} \psi, \label{e3}
\end{equation}
where $\Gamma_{\mu} \psi$ represents $\Gamma^a_{\ \mu b } \psi^b$. With respect to $\overline{\psi}$, we write
\begin{equation}
\overline{\psi}_{|\mu}=\overline{\psi}_{ ,\mu}-\overline{\psi}\Gamma_{\mu}, \label{e4}
\end{equation}
where now $\overline{\psi}\Gamma_{\mu}$ represents $\psi_b\Gamma^b_{\ \mu a}$. Notice that this identification is true only if $\gamma^{(0)}\Gamma_{\mu}^{\dagger} \gamma^{(0)}=-\Gamma_{\mu}$. There are no problems in the definition of $\psi_{a|\mu}$, it is well defined. The need of this condition is just to ensure consistency with the fact that we are assuming the spinor $\psi_a$ to be the Dirac spinor $\overline{\psi}$, which would not be true for a more general spinor. The advantage of using the matrix notation lies in reducing the number of indices. This will simplify our notation next.

\section{\label{05032011b} Dirac equation}
To define the Dirac equation in a curved spacetime, one usually uses the minimal coupling procedure (MCP for short). This approach works very well in a Riemannian geometry, i.e., there is no ambiguity in it. However, when trying to obtain a version of this equation for a spacetime with either torsion or non-metricity, one discovers that the field equations depend on where we apply the MCP. By applying it to the lagrangian (MCPL), we obtain a set of field equations that, in general, differs from the one we would have obtained by applying the MCP direct to the field equations (MCPE). Some authors have claimed that this happens because the standard variational principle cannot be used in a non-Riemannian geometry \cite{Fiziev:1995te}. Nevertheless, the most accepted approach is to assume the validity of this principle and take the MCPL as right \cite{Gasperinibook,Blagojevic:2002du,Poplawski,Fariborz,Hehl:1976kj,Hehl:1994ue,Hehl:2007bn}; that is exactly the case of the so called {\it Einstein-Cartan theory} (see e.g., chapter X of Ref.\cite{Gasperinibook}).

\subsection{The MCP}
The MCP consists basically in changing partial derivatives for covariant ones and assume that all the quantities may depend on the coordinates. Hence, a term like $\gamma^{\mu}\partial_{\mu}$ with $\gamma^{\mu}$ being constant would become $\gamma^{\mu}(x)\nabla_{\mu}$. One also substitutes the volume element $d^nx$ for  $d^nx\sqrt{-g}$.

\subsection{The MCPL}
The MCPL leads to the following Dirac action
\begin{equation}
S_D=\frac{i}{2}\int d^nx \sqrt{-g} \left[ \overline{\psi}_{|\mu} \gamma^{\mu} \psi- \overline{\psi} \gamma^{\mu}\psi_{|\mu} -2im \overline{\psi}\psi \right], \label{e1}
\end{equation}
which, in turn,  leads to the field equation\footnote{For the Dirac equation obtained by means of the MCPE, see. Ref. \cite{Adak:2002pq}.}
\begin{eqnarray}
  i\gamma^{\mu} \left[ \partial_{\mu}  +\Gamma_{\mu} + \frac{1}{2}N^{\lambda}_{\ \ [\lambda \mu]}    +\frac{1}{2}T^{ \lambda}_{\ \mu \lambda} \right]\psi  -m \psi=0, \label{e8}
\end{eqnarray}
where the torsion tensor is defined as $\mathcal{T}(V,U) \equiv \nabla_V U -\nabla_U V - [V,U]$ and its components are $T^{\alpha}_{\ \beta \nu} \equiv dx^{\alpha}\left(\mathcal{T}(\partial_{\beta},\partial_{\nu} ) \right)$. 

The presence of $N$ in both (\ref{e1}) and (\ref{e8}) is deceptive. When we make $N$ explicit in them, we find that it disappears, recall that $\Gamma_{\mu}$ depends on $N$ too. If we had applied the MCPE, the last two terms between the brackets in  Eq.~(\ref{e8}) would not be present. As we can see from (\ref{e8}), there will not be any ambiguity in the MCP for a totally antisymmetric torsion and a totally symmetric non-metricity.

\section{\label{02032011d} Final Remarks}
The covariant constancy of the Dirac matrices can be motivated by the heuristic assumption that if, in Minkowski spacetime and in Cartesian coordinates, we have $\partial_{\mu}\gamma^{\lambda}=0$, then, in a curved spactime without non-metricity, we should have $\gamma^{\lambda}_{\ \, |\mu}=0$.

The reader interested in the Dirac equation in a Riemannian manifold may consult any book on quantum gravity (see e.g., Ref. \cite{Kiefer}), or alternatively, some articles devoted to this equation in curved spacetime such as \cite{Brill:1066}. For those who are interested in non-Riemannian geometries and consider themselves beginners,  we recommend Refs. \cite{Hammond:2002rm,Gasperinibook,Puetzfeld0404119}. In Ref. \cite{Hammond:2002rm} the reader will find a review of some non-Rimannian geometries, while in Ref. \cite{Gasperinibook} the focus is on General Relativity and the Einstein-Cartan theory. Ref. \cite{Puetzfeld0404119} is a very nice review of non-Riemannian cosmology. 





%

\end{document}